\documentclass{kluwer}    

\usepackage{epsfig}

\newdisplay{guess}{Conjecture}

\newcommand{\citeP}[1]{(\citeauthor{#1} \cite*{#1})}
\newcommand{\citeN}[1]{\citeauthor{#1} (\cite*{#1})}
\newcommand{\citeNP}[1]{\citeauthor{#1} \cite*{#1}}

\newcommand{\apj} {ApJ}
\newcommand{\apjs} {ApJS}
\newcommand{\apjl} {ApJL}
\newcommand{\aap} {A\&A}
\newcommand{\solphys} {Solar Physics}
\newcommand{\arcsec} {''}

\begin{document}

\begin{article}
\begin{opening}         
\title{
SPINOR: Visible and Infrared Spectro-Polarimetry at the National Solar
Observatory} \author{Hector \surname{Socas-Navarro}, David \surname{Elmore},
    Anna \surname{Pietarila}, Anthony \surname
    {Darnell}, Bruce W. \surname{Lites}, \& Steven \surname{Tomczyk}
    \thanks{Visiting Astronomers, 
      National Solar Observatory, operated by the Association of Universities
      for Research in Astronomy, Inc. (AURA), under cooperative agreement
      with the National Science Foundation.}}

\runningauthor{Socas-Navarro et al}
\runningtitle{Visible and Infrared Spectro-Polarimetry with SPINOR}
\institute{High Altitude Observatory, NCAR \thanks{The National Center for
  Atmospheric Research (NCAR) is sponsored by the National Science
  Foundation}} 
\author{Steven Hegwer}
\institute{National Solar Observatory}
\date{Jun 28, 2005}

\begin{abstract}
SPINOR is a new spectro-polarimeter that will serve as a facility 
instrument for the Dunn Solar Telescope at the National Solar
Observatory. This instrument is capable of
achromatic polarimetry over a very broad range of wavelengths, from 430
up to 1600~nm, allowing for the simultaneous observation of several
visible and infrared spectral regions with full Stokes polarimetry. Another
key feature of the design is its flexibility to observe virtually any
combination of spectral lines, limited only by practical considerations
(e.g., the number of detectors available, space on the optical bench,
etc). 
\end{abstract}
\keywords{instrumentation: polarimeters, polarization, telescopes, Sun:
  magnetic fields,   Sun: photosphere, Sun: chromosphere}

\end{opening}           

%
%

\section{Introduction}
\label{sec:intro}

The new breed of
spectro-polarimeters scheduled to begin operation in the next
few years make the present a very exciting time for solar physics. 
Examples of these instruments are SOLIS \citeP{K98}, Solar-B \citeP{LES01},
POLIS \citeP{SBK+03},
Sunrise \citeP{SCG+03}, and the DLSP \citeP{SEL+03}. All these instruments,
however, are highly specialized and operate under fixed conditions, either
with the aim of performing synoptic observations or to optimize for spatial
resolution. While these aspects are very important for the future advance
of solar physics, we feel that there is also a need for an
``experiment-oriented'' type of instrument. By this we mean an instrument
that provides enough flexibility to implement more or less arbitrary optical
arrangements at the observer's request. Such an instrument should allow
researchers to observe any given spectral line (or combinations of lines),
either on the disk or off the limb at a diverse range of wavelengths. One
should be able to modify parameters like spatial resolution, spectral
dispersion, integration times, etc, in order to address a broad range of scientific
problems. In some sense, operating an instrument of this kind is
like setting up an experiment in a laboratory. Optical components and 
detector systems are arranged on an optical bench to investigate some
specific problem. 

For more than a decade, the highly successful Advanced Stokes Polarimeter
(ASP, \citeNP{ELT+92}) has provided these capabilities at {\it visible}
wavelengths. The 
Spectro-Polarimeter for Infrared and Optical Regions (SPINOR) will
replace the ASP as the experiment-oriented spectro-polarimeter on the
Dunn Solar Telescope (DST) at the National Solar Observatory (NSO), Sunspot,
NM, USA. SPINOR is currently under deveopment by the High Altitude
Observatory and the National Solar Observatory. 
It has been conceived with flexibility as
a top priority and intends to fill this gap in the next-generation solar
instrumentation. The new instrument will replace and enhance the capabilities
of the ASP.

Table~\ref{table1} below provides some specifications for SPINOR. The most
important enhancement over the ASP is the ability to observe simultaneously
in the visible {\it and} the near infrared. Higher spatial resolution than was
available over most of the life of the ASP is now
possible thanks the the high-order adaptive optics (AO) system \citeP{R00}
developed for the DST. Two different detectors will
be available at commissioning time (scheduled for 2006): a
PixelVision Pluto 652x488 thinned back-illuminated split frame
transfer CCD, 60 frames per second readout rate (hereafter Pluto) and a
Sarnoff CAM1M100-SFT, 1024x512, thinned back-illuminated split frame
transfer CCD, 100 frames per second readout rate.
These will be fully dedicated for 
observations between 430~nm and $\sim$1~$\mu$m. The SPINOR optics allow
for an even broader range of wavelengths, up to 1.6~$\mu$m, which will be
accessible with an HAO Rockwell infrared camera that is shared with other HAO
projects. The modular design of the instrument control allows
for additional detectors to be attached to SPINOR using NSO's Virtual Camera
interface.


This paper is organized as follows. Section~\ref{sec:science} discusses the
scientific motivations for the development of
SPINOR. Section~\ref{sec:instrument} presents a description of the
polarimeter with some details on the new achromatic optics. The polarimetric
calibration of the system consisting of the telescope and the polarimeter is
explained in section~\ref{sec:calib}. Some observations taken during two
campaigns in June 2004 and May 2005 are presented in
section~\ref{sec:obs} to illustrate the capabilities of the
instrument. Finally, section~\ref{sec:conc} presents the main 
conclusions of our work and some future perspectives.

\section{Scientific Rationale}
\label{sec:science}

The main scientific driver behind the development of SPINOR is to expand
the present ASP magnetometric capabilities into the higher solar
atmosphere. However, its infrared coverage is also 
of great relevance for the study of problems associated with photospheric
magnetism. Below we summarize some of the major scientific problems for which
SPINOR is particularly suited, which cannot be pursued using the ASP (or
other existing instruments). However, being that this an experiment-oriented
instrument designed for maximum versatility, its most important applications
might be some that we cannot envision at this time.

\subsection{Photospheric magnetism}

Recent investigations of photospheric fields in the infrared often
reveal a different picture from conventional visible observations. A
remarkable example 
is the finding of supersonic flows in the penumbrae of sunspots by
\citeN{dTIBRC01} in data from the Tenerife Infrared Polarimeter (TIP). Such
strong flows had never been observed in 
the visible, with the exception of the peculiar $\delta$-configuration
sunspots (\citeNP{MPLS+94}), and flows near pores (\citeNP{LES01}).

Perhaps the most puzzling observations of photospheric fields in the infrared
are those of the quiet sun, particularly outside the magnetic network (the
region sometimes referred to as the photospheric internetwork).
Visible
observations indicate that most of the fields in this region are strong
($\sim$1.3~kG), but concentrated into
very small areas ($\sim$1\% of the pixel; see \citeNP{SNSA02} and references
therein). However, 
recent infrared observations (\citeNP{L95}; \citeNP{LR99}; \citeNP{KCS+03})
suggest that most of the fields are weak ($\sim$400~G) and diffuse. 
These seemingly contradictory results have 
sparked a controversy on the true nature of quiet sun fields. This issue
is an important one, since our current understanding indicates that most of the
solar magnetic flux (even at solar maximum) is located in the quiet sun
outside of active regions, and it is likely that the evolution of this flux
plays a role in the heating of the upper atmosphere.
A possible solution to the observational contradiction has been proposed by
\citeN{SN03b} and 
\citeN{SNSA03}. It turns out that the observations may be explained by a
small-scale distribution of fields, beyond 
the spatial resolution of the observations, having intermixed
weak and strong fields.
Moreover, these authors showed that the actual sub-pixel
distribution of the field can be inferred from simultaneous visible and
infrared observations
(like those from SPINOR).

The results of \citeN{SNSA03} may be extrapolated to other
physical scenarios in which different field strengths coexist within the
resolution element of the observations. Another example where this
strategy would be very useful is the investigation of sunspot penumbrae. 
The actual size,
properties and origin of these filaments are still a subject of
debate (\citeNP{MP00}; \citeNP{SA01b}; \citeNP{MP01}).
The ability to infer spatially-unresolved
distributions of the magnetic field from simultaneous visible and infrared
observations would provide important clues on the structure of the penumbra.

Innovative new diagnostics of solar magnetic fields are emerging as a result
of parallel theoretical and observational advances.
In a recent effort to understand the anomalous polarization signals observed in
some spectral lines, \citeN{LATC02} studied the hyperfine
structure induced in the atomic energy levels by the nuclear spin, and its
effects on the polarization transfer process. They demonstrated that the
signature imprinted by the hyperfine structure on some spectral lines has an
important potential for magnetic field diagnostics. Some of the most
interesting lines lie in 
the wavelength domain between 800 and 1500~nm. Examples are the
Rb~I~D1 line at 794~nm , which shows a combination of hyperfine
structure and isotopic mix, or the Mn~I lines at 870~nm, and in the
infrared at 1.29, 1.33 and 1.52~$\mu$m.

\subsection{Chromospheric magnetism}

A new global picture of solar magnetism is emerging from
the seemingly disparate observational domains of photospheric small-scale 
magnetic fields and the diffuse, voluminous magnetic structure of the solar
corona.  The observation and interpretation
techniques used for photospheric and coronal studies are markedly
different. To further the development of this global view, we identify
a key missing ingredient: an in-depth investigation of the
interface layer, the chromosphere. Observational capability for chromospheric
vector magnetic fields and the associated dynamics has been lacking because most
interesting and/or useful lines
lie outside the wavelength coverage range of most solar
polarimeters (it should be noted, however, that
successful investigations of prominence fields have been carried out
recently; e.g.  \citeNP{CLA03}). Observable lines either form 
too low in the chromosphere (e.g., the Mg~I $b$-lines) or their
polarization transfer is still not well understood (e.g., H$_{\alpha}$). 

SPINOR would open new perspectives for chromospheric investigations with its
ability to observe the full polarization state of the 
Ca~II infrared triplet, around 854~nm. These
lines are the best candidates for chromospheric diagnostics, at least in the
Zeeman regime, due to their relatively simple formation physics, their long
wavelengths (which results in stronger Zeeman signals), and the
valuable information they carry on the thermal and magnetic
conditions of the higher atmosphere (\citeNP{U89}; \citeNP{SNTBRC00a};
\citeNP{SNTBRC00b}). They are also
sensitive to the Hanle effect, which 
provides complementary diagnostics on the weaker ($\sim$1~G) fields, and have
been successfully modeled by \citeN{MSTB01}.

Finally, the He~I multiplet at 1083~nm is of great interest for
chromospheric studies. This line is seen in emission in
prominences and in absorption in filaments, with strong polarization signals
arising from both Hanle and Zeeman effect. SPINOR would be able to
provide full spectro-polarimetry at 1083~nm,
which implies the potential to investigate the magnetic and
dynamic conditions of these structures. Other interesting coronal lines that
may be accessible for observations are the two Fe~XIII lines at 1074~nm,
although these may not be visible using a traditional solar telescope.

\section{Description of the instrument}
\label{sec:instrument}

SPINOR is based on the design of the ASP, which uses a rotating waveplate as
a modulator and a polarization beam splitter as a dual-beam analyzer. All of
the polarization optical components, however, have been replaced by new
achromatic ones. SPINOR utilizes the ASP calibration/modulation unit
(cal/mod) at the exit port of the DST, high-order AO, and the Horizontal
Spectrograph (HSG). 
Its optical components are distributed throughout the DST at the same
locations as those for the ASP (\citeNP{ELT+92}).
The entrance window can be covered with the achromatic linear polarizer
array. Achromatic calibration and modulation optics reside in the
calibration/modulation unit just above telescope Port 4.  The achromatic
polarizing beam splitter is just behind the entrance slit of the
horizontal  spectrograph.  Camera lenses, pre-filters, and high speed
cameras specific to SPINOR are located within the HSG. 

Since SPINOR operates over a much wider wavelength range compared to ASP,
it was deemed necessary to allow SPINOR to use a selection of gratings.
Depending upon the desired spectral line
and spectral resolution, different gratings may be
selected. Higher blaze angle gives higher spectral resolution for the same
spatial sample size. Since SPINOR will operate with AO, its design favors
a higher spatial resolution than the ASP.  As a general-purpose
research instrument, it cannot be optimized for a single line, so
the SPINOR spatial resolution is not as high as that attainable by the new DLSP.
These spectrograph issues lead typically to a spectrograph using a 40~$\mu$m
slit and 1000~mm camera lens. The observations reported in this paper were
carried out using the 308.57~line/mm grating with a blaze angle of 52$^{\rm
  o}$.

Users can choose among several NSO gratings (see
http://nsosp.nso.edu/dst/userman/instruments/hsg/) which combined with the
other variables permit a wide range of resolutions and spectral ranges. As an
example, for the Sarnoff camera, 630.2~nm, 50~$\mu$m slit, 1000~mm camera
lens focal length, 316~line/mm grating, and 63~degree angle of incidence,
resolving power is 170,000, and spectrum length is 1.2~nm.

\subsection{Specifications}

SPINOR has been conceived keeping versatility as the highest priority in
order to allow for a broad range of potential applications. The project is
being developed in a way that allows the ASP to remain operational until the
new instrument is completed. Its most important features are:

\begin{table*}
\caption[]{Performance comparison between SPINOR and ASP}
\label{table1}
\begin{tabular}{lccc}
\hline
Parameter &  & ASP & SPINOR \\
\hline
Calibratable wavelength range (nm) &  & 450-750 & 430-1600 \\
Field of view along slit (arc seconds) &  & 80 & 120 \\
Quantum efficiency & 430 nm & 0.01 & 0.72 \\
                   & 700 nm & 0.32 & 0.80 \\
                   & 1080 nm & 0.00 & 0.03, 0.60$^{*}$ \\
                   & 1600 nm & 0.00 & 0.60$^{*}$\\
                   &            &      & $^{*}$(with suitable IR camera)\\
\hline
\end{tabular}
\end{table*}

\begin{itemize}
\item Achromatic optics from 430 to 1600~nm offer the
  capability of simultaneous observations at diverse wavelengths
  This extended range represents an important
  improvement over that of the ASP. Its potential scientific
  benefits have been discussed in more detail in section~\ref{sec:science}.
\item Detectors with higher quantum efficiency than those
  in ASP allowing for higher signal to noise observations.
\item Use of state-of-the-art components, electronic systems, computers,
  detectors and software. Deployed in 1991, ASP's technology is now
  15 or more years old, and some of its systems are starting to fail. Both routine 
  maintenance tasks and normal operation are compromised because many
  replacement parts are no longer available. The SPINOR system 
  will be more stable and suffer significantly less
  downtime. Data products will also be easier to handle 
  (e.g., using DVDs instead of magnetic tapes, simpler
  analysis procedures, etc), an improvement that will greatly facilitate
  the science and make the instrument more accessible to a broader user
  community.
\item Open and modular optical design, with room on the optical bench to incorporate
  and/or 
  replace components. SPINOR will be deployed at the DST as
  a set of instrument modules with ``virtual'' cameras, a concept that NSO
  has developed for the DLSP and other DST instrumentation. 
  This will allow diverse and complex observations,
  combining SPINOR with other DST instruments. The control software will be
  fully 
  customizable for a broad variety
  of observing modes (although several pre-defined modes will exist for
  frequently used configurations).
\end{itemize}

\subsection{Achromatic polarization optics}
\label{sec:optics}

\begin{figure*}
\includegraphics[width=1.\maxfloatwidth]{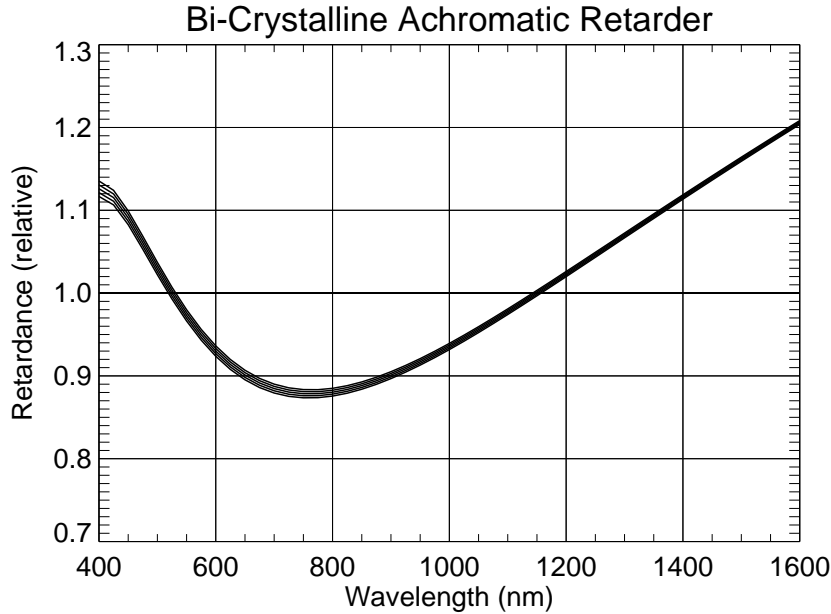}
\caption{
Design performance for the SPINOR retarders as a function of wavelength.
There are two 
curves shown. The upper one is for 20~C and the lower one for 30~C.
These are shown for a one wave of retardance. 
\label{fig:opti1}
}
\end{figure*}

Fig~\ref{fig:opti1} shows the retardation as a function of wavelength for the
SPINOR bicrystalline achromatic retarders (Meadowlark Optics, see
\citeNP{GE04}). Plates of different retardations are obtained by varying the
thicknesses of the quartz and sapphire parts (maintaining the thickness ratio).
SPINOR uses a 1/4 wave retarder for polarization calibration, a 0.375 wave
retarder for polarization modulation, and a 0.50 wave retarder to perform a
coordinate transformation between the slit and the polarizing beam splitter.
Crystals in the calibration retarder are broad-band anti-reflection (AR)
coated and air spaced using epoxy with glass beads imbedded.  One side of the
50~mm calibration polarizer uses 30~$\mu$m beads, the other side 10~$\mu$
beads.  The 50~mm diameter polarization modulator consists of, in order, a
broad band AR coating, sapphire, CeF$_3$ coating, adhesive, quartz, and a
broad band AR coating.  The 25~mm diameter half-wave retarder is constructed
like the modulator.  The broad band AR coating (from ZC\&R) has a reflectance
typically $<$1.5\% from 450~nm to 1600~nm.

\begin{figure*}
\includegraphics[width=1.1\maxfloatwidth]{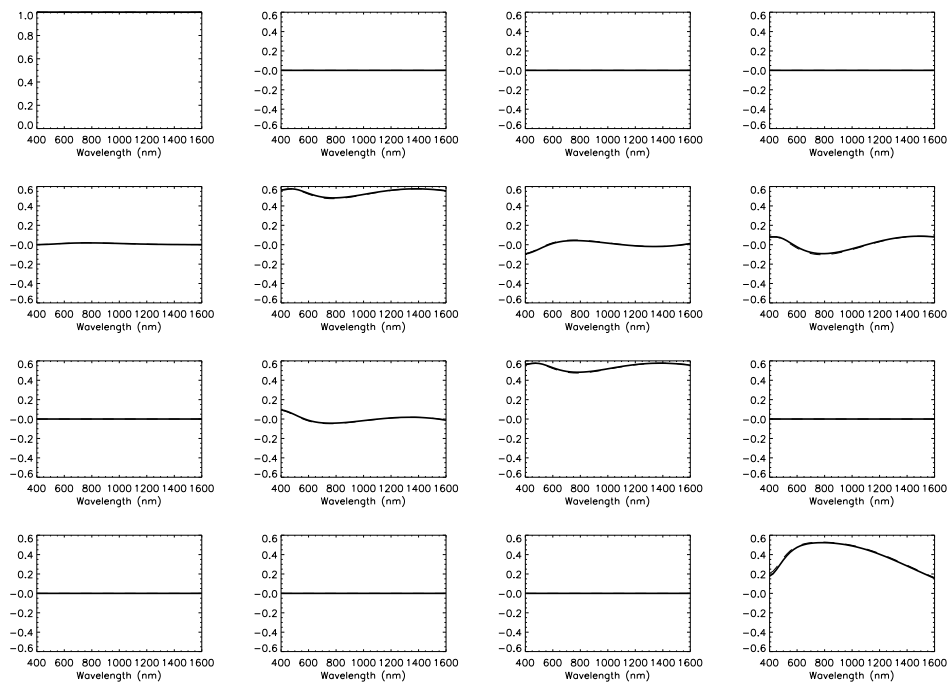}
\caption{
Theoretical polarimeter response matrix elements as a function of
wavelength. Each plot represents a matrix element.
The solid and dashed curves, which virtually overlap at the scale of the
figure, represent different temperatures of 30 and 20 degrees Celsius,
respectively. 
\label{fig:opti2}
}
\end{figure*}

Fig~\ref{fig:opti2} shows the theoretical polarimeter response matrix
as a function of wavelength 
for SPINOR. This is calculated using the retardance 
design curves for the bicrystalline achromatic modulator and
half-wave retarder behind the slit.

SPINOR uses wire grid polarizers (Versalight) for the telescope entrance
window polarizers, calibration polarizer, and in the polarizing beam
splitter. These polarizers transmit one orientation of linear polarization
and reflect the orthogonal polarization.
Versalight aluminum wire grid polarizers produce an extinction ratio of 400
at 450~nm and 4000 at 1650~nm.  There is a dip at 950~nm to just under
1000. The wire grid side of the polarizer can easily be damaged, therefore
entrance window polairzers have the wire grid surface protected by a MgF$_2$
coating.  The calibration linear polarizer has custom broad band AR coatings
optimized for the wire grid and glass surfaces.  The polarizing beam splitter
cube is a sandwich of two 90-degree prisms with the polarizer in between,
cemented with an index matching adhesive.


\subsection{Detectors}
\label{sec:detec}

SPINOR uses the HSG at the DST, which sits on top of an optical bench and
allows for great flexibility in the instrumental arrangement.  The data
presented here have been collected using a Rockwell TCM 8600, Pixel Vision
Pluto, and the ASP TI TC245 cameras. SPINOR also has a dedicated Sarnoff
CAM1M100.  Any high frame rate camera that reads out upon an external strobe
that is connected to a computer capable of storing the images according to
the modulator status as received on a printer port can be connected to
SPINOR. Camera computers receive modulator position information on their
parallel ports, which is used to phase the camera reads to the nearest 1/16
of the modulator rotation. The following table  
lists properties of current and possible SPINOR 
cameras.  The SWIR camera listed is similar to cameras produced by Sensors
Unlimited and Indigo.

\begin{table}
\begin{tabular}{cccccc}
\hline
Wavelength    & Sarnoff &  PixelVision  &  ASP     &  Rockwell  &  SWIR \\
(nm)          & CAM1M100 &     Pluto or BioXight & TI TC245 & TCM8600   &       \\
\hline
450	&	.73	&	.73	&	.20  &	-	&	- \\
650	&	.78	&	.85	&	.42  &	-	&	- \\
850	&	.59	&	.46	&	.30  &	-	&	- \\
1100	&	(.1)	&	(.05)	&	(.02) &	$>$0.50	 &	0.65\\
1600	&	-	&	-	&	-   &	$>$0.50	 &	0.70\\
\hline
Max Frame &             &               &           &           &          \\
Rate (fps)&	100*	&	50*	&	60&	30	&	$>$29*\\
Read Noise &   &   &   &   &  \\
electrons/read&	40 &		50	&	50&	70	&	$\sim$300+\\
Format	&	512x1024	&488x652	&	230x512 &
1024x1024	& 512x640 \\
Pixel size ($\mu$m)		&16x16&		12x12&		8.5x19.75& 18x18	&
25x25 \\
\hline
\end{tabular}
\caption[]{Quantum efficiency (as a function of wavelength), and other
  relevant parameters of various cameras for SPINOR.}
\label{table:detec}
\end{table}

Quantum efficiency values are vendor specifications and values in parenthesis
are extrapolated.  Frame rates with an asterisk can be increased with reduced
field of view.  Read noise values are measured except for the SWIR camera,
which is a vendor specification.

\begin{figure*}
\includegraphics[width=1.5\maxfloatwidth]{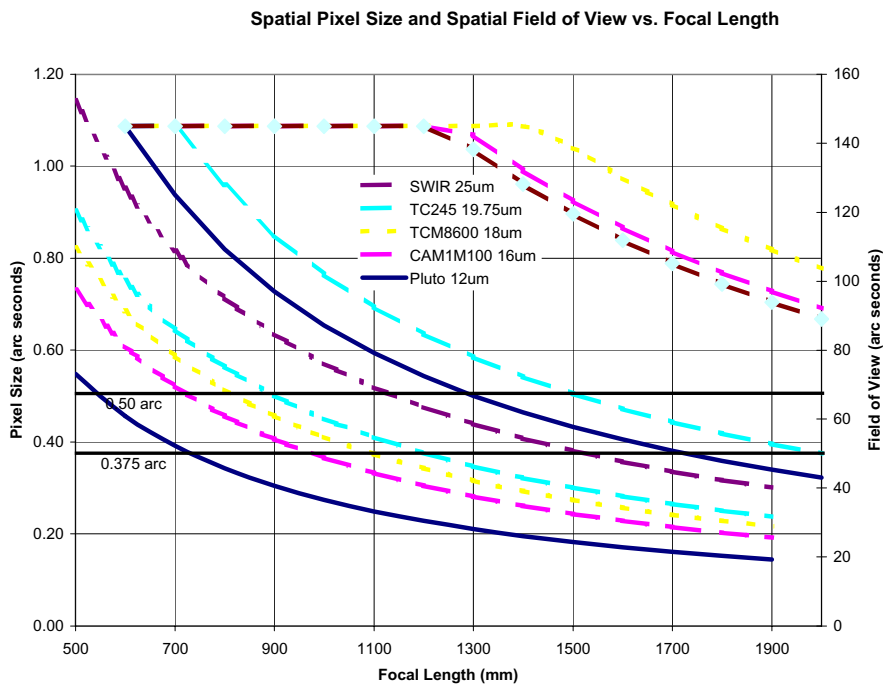}
\caption{ Pixel size and field of view for various cameras and camera
lenses.The lower set of curves show pixel size according to the scale on the
left. The upper curves show field of view as shown on the scale on the
right.
\label{fig:pixels}
}
\end{figure*}

Spatial pixel size and field of view vary depending upon the focal length of
the spectrograph camera lens used.  Figure~\ref{fig:pixels} shows pixel
spatial sample size and field of view for various cameras.  This plot assumes
the DST HSG with standard 3040~mm focal length collimator
lens and the standard f/36 beam feeding the spectrograph, which produces a
133~$\mu$m/arc-second plate scale at the HSG entrance slit.  Typical SPINOR
slit widths are 50~$\mu$m (0.375'') or 80~$\mu$m (0.60'').  
In computing the
field of view, pixel size, camera lens focal length, and the full field of
view of the polarizing beam splitter have been considered.  The polarizing
beam splitter has a field of 145'' with a 30'' separation between the two
beams.

\subsection{Calibration and characterization}
\label{sec:calib}

The calibration procedure for SPINOR consists of the determination of the
Mueller matrix of the telescope (${\bf T}$) and the polarimeter reponse
(${\bf   X}$). Any solar Stokes vector ${\bf s_{in}}$ is observed as ${\bf
  s_{out}}$ (which is not necessarily a Stokes vector): 
\begin{equation}
{\bf s_{out}} = {\bf XT s_{in}} \, .
\end{equation}

The polarimeter response matrix depends on the particular instrumental
configuration and varies significantly between observing runs (sometimes
even from one day to the next). For this reason, polarimeter calibration
operations are done typically once a day, or even more often if changes are
made to the instrument. Fortunately, calibration operations do not take too long
($\sim$20 minutes). A calibration linear polarizer and retarder are mounted
following the telescope exit window. These are used to polarize the light
beam in various known states. The known input vectors and the observed
outputs are used in a least-squares fitting procedure 
to determine the matrix elements of ${\bf X}$. This is essentially the same
procedure that was used for the ASP calibration, which is explained 
in the paper by \citeN{SLMP+97} and will not be repeated here. The only
difference is that the calibration
polarizer and retarder are now almost entirely achromatic over the broad
wavelength range of SPINOR. The retardance of the calibration retarder and
the orientation of its fast axis are taken as 
free parameters of the fit, in order to account for wavelength
variations and possible mounting inaccuracies. The following are ${\bf
  X}$-matrices measured in the May 2005 run at 587.6, 1083.0 and 1565.0~nm:

\begin{eqnarray}
\label{X1}
{\bf X}_{587.6} = \left ( \begin{array}{cccc}
      1.00  &   0.24 &  4.08\times10^{-4} &  -6.16\times10^{-2} \\
    1.79\times10^{-2}  &   0.46 &     0.15 &    2.99\times10^{-2} \\
  -4.11\times10^{-3}  &  -0.14 &     0.44 &     0.13 \\
   1.84\times10^{-3}  &  3.41\times10^{-2} &    0.15  &  -0.46
                  \end{array} \right )
\end{eqnarray}

\begin{eqnarray}
\label{X1}
{\bf X}_{1083.0} = \left ( \begin{array}{cccc}
     1.00 &    0.23  &  4.90\times10^{-2} &  -6.99\times10^{-2} \\
    1.17\times10^{-2} &     0.45 &     0.18 &    2.47\times10^{-2} \\
  -3.64\times^{-2} &   -0.18 &     0.45 &    8.19\times10^{-2} \\
   1.69\times10^{-2} &    4.87\times10^{-2} &    9.91\times10^{-2} &    -0.41 
                  \end{array} \right )
\end{eqnarray}

\begin{eqnarray}
\label{X1}
{\bf X}_{1565.0} = \left ( \begin{array}{cccc}
      1.00  & 6.22\times10^{-2} & 2.95\times10^{-4} &   -1.38\times10^{-2} \\
  6.29\times10^{-3} &   0.43 &     -0.41 &    -4.99\times10^{-2} \\
  -4.25\times10^{-4} &   0.44  &  0.43 &  4.08\times10^{-2} \\
  5.02\times10^{-2} &  3.84\times10^{-3} & 2.59\times10^{-2} & -0.19
                  \end{array} \right )
\end{eqnarray}

The telescope calibration requires an entire day of observations at different
positions of the Sun on the sky. This needs to be done at least every time
the turret mirrors are recoated, and desirably every few months.
We have devised a new method that augments that of
\citeN{SLMP+97}. In that paper, the
telescope is modeled as a function of the following 
free parameters: the entrance window retardance and axis orientation; the
turret mirrors retardance and ratio of reflectivies ($rs/rp$); the main
mirror retardance and $rs/rp$; the exit window retardance and axis
orientation; and finally the relative rotation between the telescope and the
polarimeter (see \citeNP{SLMP+97} for an explanation of these
parameters). 

An array of achromatic linear polarizers, placed in front of 
the entrance window, is used to introduce known states of polarization in the
system. These vectors are measured at various times of the day, covering a
wide range 
of the three variable angles of the telescope: turret azimuth, elevation
and table rotation. Again, a least-squares fit is used to determine the free
parameters of the model.

\begin{figure*}
\includegraphics[width=1.2\maxfloatwidth]{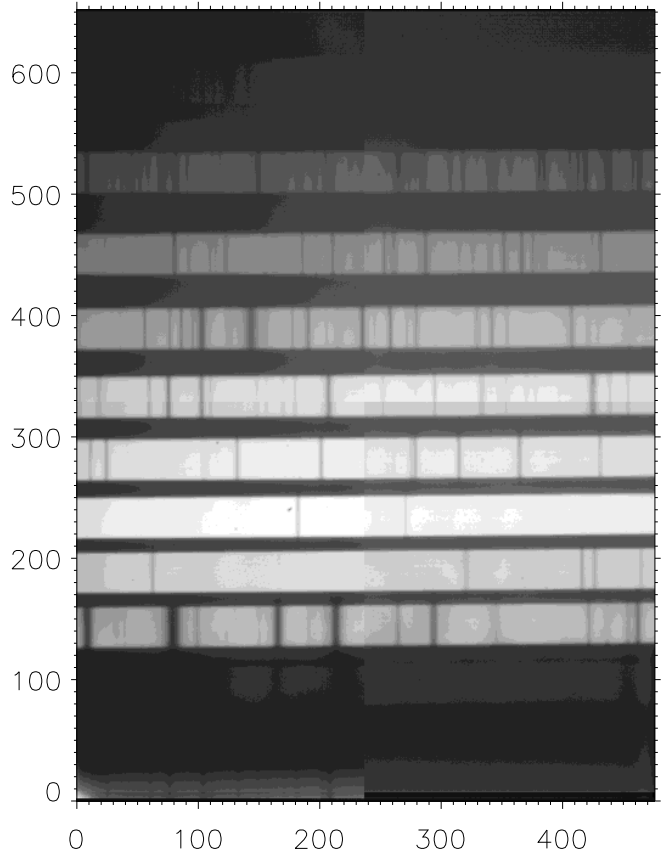}
\caption{
Cross-dispersed spectra from a telescope calibration operation. Central
wavelengths in nm from top to bottom: 416, 451, 492, 541, 601, 677, 773,
902. }
\label{fig:tels}
\end{figure*}

\begin{figure*}
\includegraphics[width=1\maxfloatwidth,height=18cm]{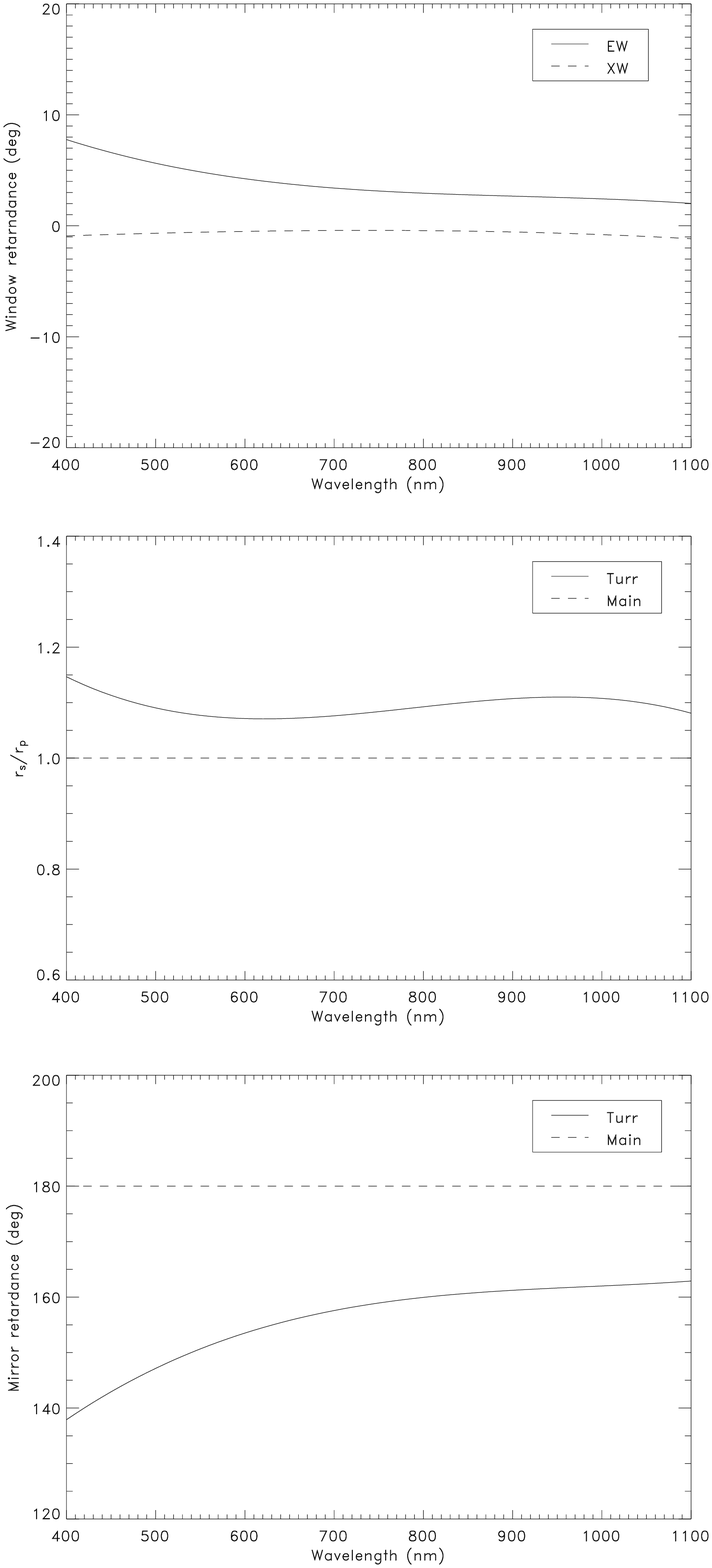}
\caption{
\label{fig:Tmat}
Curves of retardance (degrees) and $r_s/r_p$ as a function of wavelength,
determined from the telescope calibration operation using the cross-dispersed
spectra.
}
\end{figure*}

The new SPINOR technique uses a cross-dispersing prism placed before one
of the detectors to record the  
various overlapping orders simultaneously (shown in Fig~\ref{fig:tels}). In
this manner we are able to 
obtain telescope calibration data for many wavelengths at once. Instead of
fitting each set of parameters independently for each wavelength, we take a
fully-consistent approach whereby the entire dataset is fit using a
wavelength-dependent model. Some parameters (window
retardances and mirror properties) are allowed to vary with
wavelength in the fit, up to the third order in $\lambda$. This is enough to
account for the non-monotonic behavior in the complex refractive index of
aluminum in the near infrared. The rest of the parameters 
are wavelength-independent. This model has sufficient freedom to reproduce
the measured vectors in the entire dataset and at the same time provides a
consistent model of the telescope. Table~\ref{table:tel} shows the various
telescope parameters obtained from our calibration at three different
wavelengths. Fig~\ref{fig:Tmat} shows the wavelength dependence of the
retardance and $r_s/r_p$ of the telescope mirrors and windows. This
calibration has been successfully applied on actual data at 587.6, 630.2,
849.8, 854.2, 866.2, 1083.0 and 1565.0~nm, which demonstrates the robustness
of the method. 

\begin{table} %
\begin{tabular}{lccc}                                        
\hline
Wavelength (nm)  &  400   &   600   &   800  \\
\hline
EW retardance  &  7.53   &  3.79   &  2.67   \\
EW orientation &   38.64   & 38.64    &  38.64     \\
Turr. rs/rp    &  1.13  & 1.07    & 1.10  \\
Turr. retardance & 139.11 & 155.62  & 161.23  \\
Main rs/rp     &  1.00  & 1.00    & 1.00   \\
Main retardance     &  179.99  & 179.99    & 179.99   \\
XW retardance  &  1.56  &   0.97   &   0.28   \\
XW orientation  &  345.81   &  345.81   &   345.81  \\
T-X rotation   &  81.26   &  81.26   &  81.26  \\
\hline
\end{tabular}
\caption[]{DST telescope parameters retrieved with our calibration procedure.
EW: Entrance window, Turr: turret mirrors, XW: Exit window. All angles are
degrees. 
}\label{table:tel}
\end{table}

\section{Observations}
\label{sec:obs}

A first engineering run took place on June 2004 with the following
configuration. The new polarization and imaging optics were carefully set up
and aligned. For much of this run we used the old ASP modulator because at
the time of the observations we had not yet received the new achromatic
modulator from the manufacturer. This resulted in less than optimum
polarimetric efficiency at long wavelengths. The new modulator arrived later
in the observing run and is now available for use at the DST. The ASP cameras
were configured to record the Ca~II lines at 849.8 and 854.2~nm (plus some
other photospheric and telluric lines blended in the wings of the Ca
lines). The new Pluto camera was observing at 1083.0~nm, including the
He~I multiplet, two photospheric lines and a telluric line. We aligned the
camera chips so that the broad Ca lines would fall near the edge of the
chip. In 
this manner we can record some wing ``continuum'' on the opposite side of the
detector, which is useful for the data analysis (polarization
calibration, flat-fielding and also as a reference for the thermodynamics in
the inversions). Since most of the 
polarization signal is concentrated near the line core, it is not necessary to
record both wings of the Ca lines. An additional advantage is that we can
also record a strong photospheric line in the 854.2~nm images.

\begin{figure*}
\includegraphics[width=1.2\maxfloatwidth]{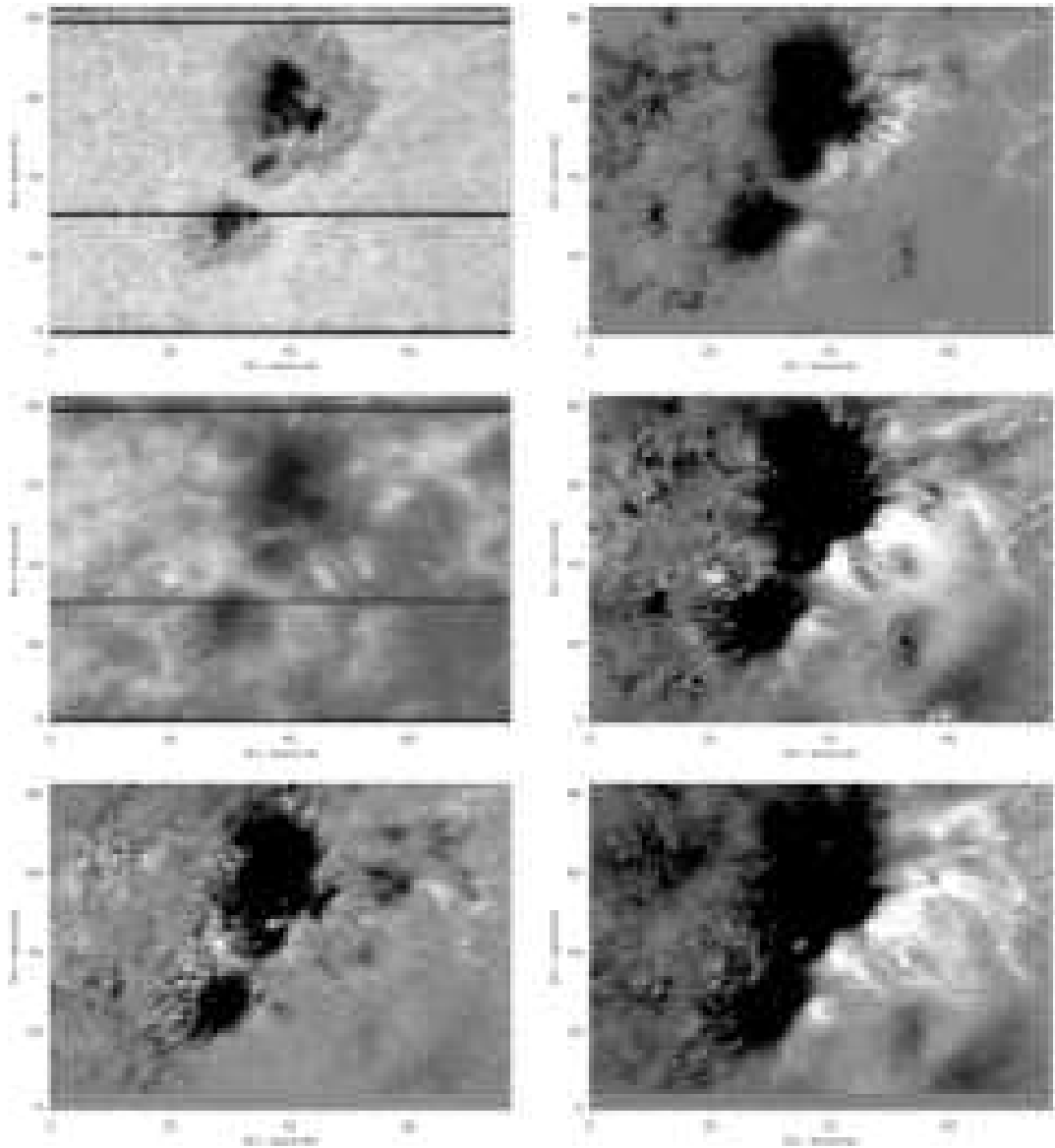}
\caption{
Various maps of active region NOAA~0634. From top to bottom, left to right:
Continuum intensity at 854~nm. Photospheric magnetogram. Chromospheric
filtergram in 
the core of the Ca~II 854.2~nm line. Chromospheric magnetogram taken on
the red side of the 849.8~nm line core. Chromospheric magnetogram on the
blue side of 854.2~nm. Chromospheric magnetogram on the red side of 854.2~nm
. The gray scale in all chromospheric [photospheric] magnetograms is $\pm$2\% 
[$\pm$3\%] of the corresponding Stokes I intensity (see text). Disk center is
towards the left of the image.
}
\label{fig:maps}
\end{figure*}

Fig~\ref{fig:maps} shows several maps of active region
NOAA~0634 observed on June 16 at 15:16~UT. This dataset has the best seeing
among our active region observations, with a 
granulation contrast of 3.3\% (notice that this value may not be directly
comparable to that of visible observations because of the variation of the
source function with wavelength). Some features apparent in the
magnetograms 
or the chromospheric filtegrams (e.g., the transversal fluctuations in the
penumbral filaments near coordinates [$x=50$,$y=55$] in the figure) exhibit
spatial scales as small as 0.6\arcsec. 

The maps represented in the figures have been obtained from a
spectro-polarimetric scan of the region consisting of 350 steps of
0.22\arcsec each. The 
scanning step oversamples the resolution element. This was done so that it
would be possible to bin the 1083.0~nm data in order to build up its photon
count 
without saturating the other two detectors. Unfortunately the 1083~nm
data from this day was unusable due to problems with the camera. Therefore we
only show the 849.8 and 854.2~nm regions for this dataset.  

\begin{figure*}
\includegraphics[width=1.2\maxfloatwidth]{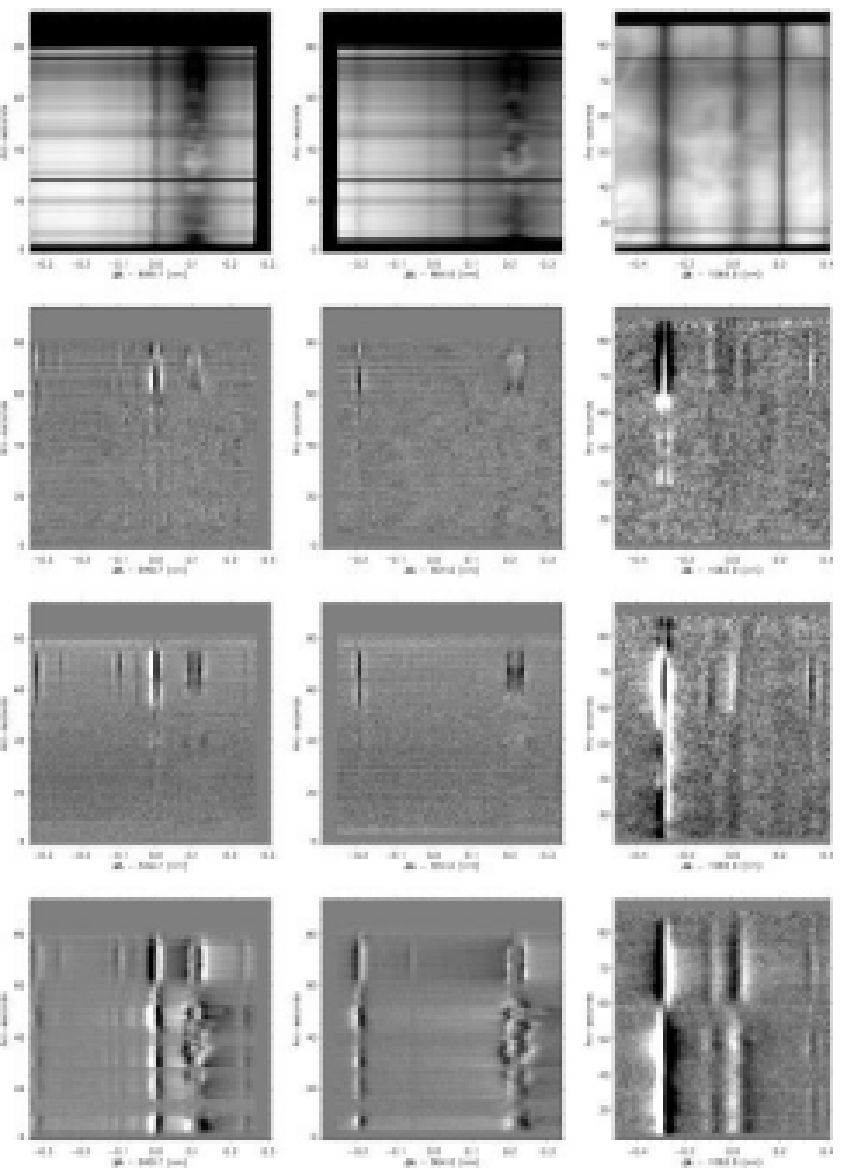}
\caption{
Stokes profiles from one particular slit position of the NOAA~0635
scan. Rows: Stokes I, Q, U and V parameters (from top to bottom). Columns:
Images from the camaras at 849.8 (left), 854.2 (middle) and 1083 (right) nm.
Profiles are normalized to quiet Sun continuum. Stokes Q and U panels are
saturated at $\pm$1\%, Stokes V panels are saturated at $\pm$2\%.
}
\label{fig:profs}
\end{figure*}

The term ``magnetogram'' is used in this work somewhat inappropriately (but
following a very widespread terminology) to refer to the amount of
Stokes V signal integrated over a certain bandwidth on either the red or blue
side of the line center, and normalized to the intensity integrated over the
same bandwidth. We have used a 10-pixel wide square filter
function. Notice that this quantity is not necessarily related to the
magnetic field strength in the Sun. In some cases it may bear some
resemblance to the longitudinal magnetic flux density (as defined by
\citeNP{GLASN+02}), but this is not always the case (\citeNP{SN02})
especially given the complex polarization patterns exhibitted by the
chromospheric lines (see Fig~\ref{fig:profs}). For this reason, the red
and blue magnetograms represented in Fig~\ref{fig:maps} are different. 

A sample of the slit spectra recorded by SPINOR is presented in
Fig~\ref{fig:profs}. The data in the figure are from a scan of NOAA~0635
observed on June~19 at 14:08~UT. The problems with the Pluto camera computer
had been 
solved during the previous days, allowing us to record data at 1083~nm.
The signal-to-noise ratio 
is quite poor due to 1) the low polarimetric efficiency discussed above,
and 2) especially to the low quantum efficiency of the camera at this
wavelength ($\simeq$3\%) (these issues were addressed in the subsequent May
2005 run by using the new achromatic modulator and the Rockwell
detector). The 1083~nm data displayed in the figure have been binned in the
scanning direction over a 1.5\arcsec \, interval. The 849.8 and 854.2~nm data
are unbinned (0.22\arcsec).

\begin{figure*}
\includegraphics[width=1.2\maxfloatwidth]{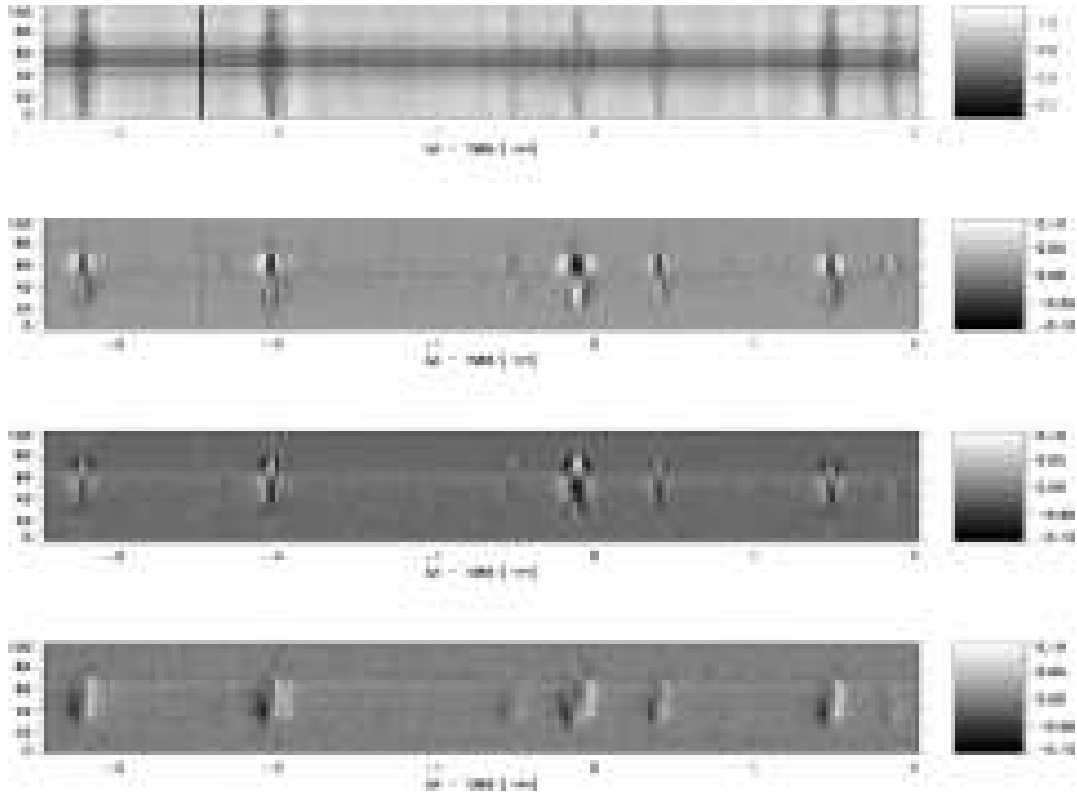}
\caption{
Stokes profiles from one particular slit position of a sunspot map observed
on 23 May 2005. Rows: Stokes I, Q, U and V parameters (from top to
bottom). Profiles are normalized to quiet Sun continuum. Wavelengths are in
nm from 1565.
}
\label{fig:profsIR}
\end{figure*}

A second observing campaign was recently conducted on May 2005, intended to
characterize and demonstrate the capabilities of SPINOR beyond 1~$\mu$m with
the infrared Rockwell camera. In this campaign we carried out observations up
to 1565~nm. A sample of such observations after our preliminary analysis is
presented in Fig~\ref{fig:profsIR}.  

\section{Conclusions}
\label{sec:conc}

This paper presents a detailed description and some initial results from
first-light observations with SPINOR, a new facility instrument for the DST.
We perceive an urgent demand in the solar community for a new
experiment-oriented (as opposed to specialized) spectro-polarimeter. We
expect that SPINOR will fulfill this need and remain at the cutting-edge of
solar research at least until the construction of the Advanced Technology
Solar Telescope (\citeNP{KRH+02}). Its broad wavelength coverage will provide
a uniquely connected view of photospheric and chromospheric
magnetism. The observations presented in the present paper are
intended to emphasize this point and to demonstrate the capabilities of the
instrument in the near infrared.


Future work will focus on the implementation of a new control system. SPINOR
is currently being operated from the ASP control computer, which sends strobe
signals to both ASP cameras and up to four other ``external'' detectors. The
new control system, which is expected to be implemented by the end of 2006,
will be more user-friendly and take advantage of the virtual camera interface
currently under development by NSO for the DST.

As with the ASP, researchers from national and foreign institutions 
will be able to access the new instrument through the usual time allocation
competition for the Dunn Solar Telescope, operated by NSO.

\section{Acknowledgments}

The authors wish to acknowledge the enthusiastic support from the NSO staff
at the Sacramento Peak observatory, especially D. Gilliam, M. Bradford and J.
Elrod. Thanks are also due to S. Gregory, R. Dunbar, T. Spence,
S. Fletcher, C. Berst and W. Jones.


\end{article}

\end{document}